\documentclass[10pt,a4paper]{article}
\usepackage[utf8]{inputenc}
\usepackage[T1]{fontenc}
\usepackage{bm,amsmath,amssymb,mathtools,color,graphicx}
\usepackage{textcomp,gensymb,units,lipsum}
\usepackage[sc]{mathpazo}
\usepackage{layaureo}
\usepackage[pdftex]{hyperref}

\graphicspath{{Figures/}}

\newcommand{\inlimine}[2]{\begin{flushright}\begin{minipage}{0.4\textwidth}\textit{#1}\begin{flushright}\textsc{#2}\end{flushright}\end{minipage}\end{flushright}\vspace*{0.5cm}}

\title{Evolution and Probability\footnote{\copyright Luca Peliti 2019}}
\author{Luca Peliti\\SMRI (Italy)\\
\texttt{luca@peliti.org}}
\date{January 6, 2019} 

\begin{document}
\maketitle

\thispagestyle{empty}

\begin{abstract}
Life forms exhibit such a degree of exquisite organization that it seems impossible that they could have developed out of a process of trial and error, as intimated by the theory of Darwinian evolution. In this general public paper I discuss how differential reproduction rates work in producing an exceedingly high degree of improbability, and the conceptual tools of the theory of evolution help us to predict, to some degree, the course of evolution---as it is routinely done, e.g., in the process leading to the yearly influenza vaccines.
\end{abstract}

\inlimine{Natural selection is a mechanism for generating an exceedingly high degree of improbability.}{R. A. Fisher}

\subsubsection*{A priori vs.\ a posteriori probabilities.} 
\begin{figure}[htb]
\begin{center}
\includegraphics[width=\textwidth]{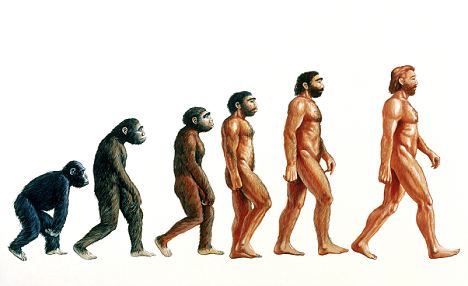}\\
\end{center}
\caption{A popular representation of the steps of human evolution. From the \textsl{Daily Mail}.}
\label{fig:evoHuman}
\end{figure}

\begin{figure}[htb]
\begin{center}
\includegraphics[width=\textwidth]{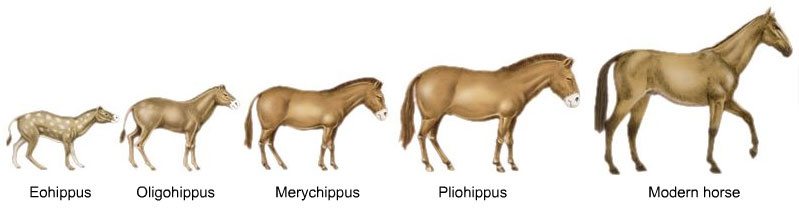}
\end{center}
\caption{A representation of the evolution of the horse. From \textsl{Pinterest}.}
\label{fig:evoHorse}
\end{figure}

Figure~\ref{fig:evoHuman} shows a popular image of the process of human evolution. We see, starting from a quite ape-like ancestor, a sequence of life forms that progressively become more similar to a present idea of a human, of course male, light-skinned, fair-haired, probably anglo-saxon.
Even when one considers other life forms, one gets a similar picture, as in the case of the evolution of the horse (fig.~\ref{fig:evoHorse}). In these representations he evolutionary process appears as the progressive discovery of an archetype (\textsl{Homo sapiens} or \textsl{Equus caballus}) which existed previously, at least in the world of ideas.

\begin{figure}[htb]
\begin{center}
\includegraphics[width=\textwidth]{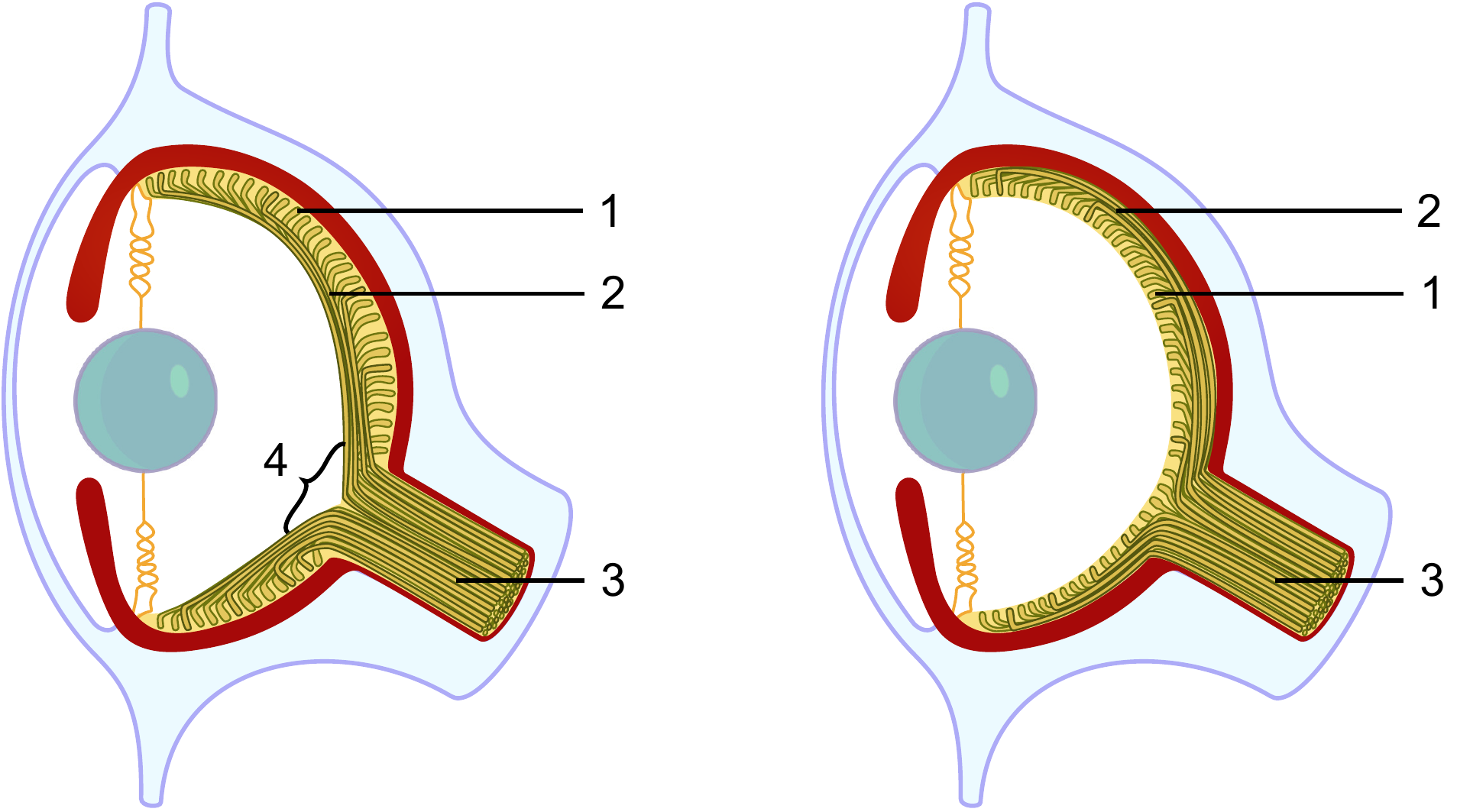}
\end{center}
\caption{Left panel: Scheme of the structure of the vertebrate eye. Right panel: Scheme of the structure of the octopus eye. 1: Retina, 2: Nerve fibers, 3: Optical nerve, 4: Blind spot. From \textsl{Wikipedia}.}
\label{fig:evoEye}
\end{figure}

If we accept this image, the theory of evolution faces a great problem: How can this archetype be reached in a reasonable time by the Darwinian mechanism of selection and mutation? To make one example: How can we think that such a complex structure as the human eye (on the left in Figure~\ref{fig:evoEye}) could be formed by a trial-and-error process? In fact, one can see that the formation of eyes in vertebrates is not the only phenomenon of its kind. During the evolution of Life on Earth, other life forms have developed comparable structures. One of these life forms is the common octopus. Looking at the details of the anatomy of the octopus eye (on the right in Figure~\ref{fig:evoEye}), we see that, although its structure is quite close to that of the vertebrate eye, it is the outcome of a radically different process. For example, the retina (1) in the vertebrate eye is made of cells connected to the brain via nerve fibers (2) which gather to form the optical nerve (3): the nerve fibers are placed \emph{above} the retina, so that the light must cross them in order to reach the light-sensitive cells. Indeed, the point where the optical nerve crosses the retina to enter the eye corresponds to the blind spot (4). The octopus eye seems to have benefitted of a more rational design: the nerve fibers (2) lie below the retina (1), and therefore there is no blind spot, and the connections with the remainder of the nervous system take place in an optical ganglion (not represented) placed behind the eye globe. 

We can try and make these ideas more precise by considering the evolution of a typical protein. Let us follow an argument by Hoyle and Wickramasinghe~\cite[cap.~2]{Hoyle}, who try to prove that the  usual evolution theory based on the mutation-selection mechanism cannot explain the formation of complex biomolecules from a random association of atoms. Following their line of argument, let us evaluate the probability that placing amino acids at random in a linear sequence we obtain a working enzyme. Living cells have a repertoire of 20 amino acids and proteins are made of linear chains of amino acids (called polypeptide chains). A typical enzyme (a kind of protein whose function is to fasten, sometimes by several orders of magnitude, the unfolding of chemical reactions) is about a hundred amino acids long. If we suppose that only one particular amino acid sequence can make a functional enzyme, the probability to obtain this sequence by randomly placing amino acids will be of the order of one over the total number of amino acid sequences of length 100, i.e.,  $20^{100}\approx 10^{30}$. Of course this is a very pessimistic estimate, since it supposes that placing even only one wrong amino acid makes the sequence non-functional. But a more realistic calculation does not reach substantially different conclusions~\cite[p.~24]{Hoyle}. If this is the case, it will be hard to escape Hoyle and Wickramasinghe's conclusion that life on Earth got started by structures of extra-terrestrial origin (whatever it may be) which had already solved the basic problems of biochemistry. And therefore that the passage from non-living to the living started somewhere else in the vast extension of the Universe, or even outside it.

We can recast this argument in the following way. Let us suppose, following Dawkins~\cite[Chap.~11]{Dawkins}, that 1000 evolution steps are necessary to obtain a working eye, and that at each step there are only two possibilities: the right or the wrong one. If the choice is made blindly, the probability of making the right choice at any step is equal to $\frac12$. Then the probability of always making the right choice at every step is equal to $1/2^{1000}\approx 1/10^{300}$. It is clear that the fact that our lineage made the correct choice at each step looks nothing less than miraculous!

However, neither the eye nor the protein need to be perfect---in the sense of possessing their present form---in order to provide advantage to the organisms that produce them. It is unrealistic, therefore, to evaluate the probability of producing them in the way that we just described. The eye is the result of a ``Research and Development'' process made by many life forms during long stretches of time, starting from the formation of small photosensitive patches, that came to be later organized in more complex and effective structures. And the more effective structures were retained because they provided advantages to the life forms that carried them---advantages that materialized as a more numerous offspring.

In the argument we just discussed we assumed that one does not know if one is on the right path until the end is reached. Let us suppose instead that each step made in the good direction provides a small advantage in terms of survival or fecundity to the being that makes it. More precisely, let us imagine to send in this ``Garden of forking paths'' a group of, say, 100 individuals, who perform their choice (right or wrong) at each step, and then reproduce.  Let us assume that those who make the right choice at the right moment have slightly more offspring than those who make the wrong one: for instance, that for each offspring of an individual who made the wrong choice, there are 1.02 offspring (on average) of one who made the right choice. Thus, after the first step, we shall have 50 individuals on the right side and 50 on the wrong side and, after reproduction, 102 on the right side and 100 on the wrong side. This is surely a very small difference: the fraction of individuals on the right path is 51\%\ rather than 50\%. However, if we wait a few generations, the fraction of individuals on the right side will increase: after 10 generations the ratio of the number of offspring of an individual who had made the right choice to that of one who made the wrong choice will be about 1.22, and after 100 generations it will be about 7.25. Thus after 100 generations about 88\%\ of the population will be on the right path. Imagine that at this point those who are on the right path face a new bifurcation. Half of them will take the right side, and half the wrong one. Thus we shall have 44\% of individuals on the right path, 44\% who made the wrong choice at the last fork, and 12\% of those who had made the wrong choice at the beginning. After waiting 100 generations more, each individual who made the right choice twice in a row will have (on average) 7.25 more offspring than those who made the wrong choice at the last step, and each of them will have (on average) 7.25 more offspring than those who got their first step wrong. Summing up, there will be 88\% of the population on the right path, a bit less than 12\% having got the last step wrong, and less than 1\% of those who got their first step wrong. Continuing this way, and assuming to be able to wait for 100 generations at each step, in $100\,000$ generations (a long time, but reasonable on the geological scale) our population will be made for 88\% of individuals whose ancestors made all the correct choices, about 12\% of those who got only the last step wrong, and the offspring of all the others will make a completely negligible fraction of the population. Looking at this path in retrospect it appears as extremely improbable, since its a priori probability is extremely small. However, the population took advantage at each step of the hints of the selection, so that its a posteriori probability is large enough. Moreover, it may well be that our guess on the a priori probability is too pessimistic, for two reasons: one, that it is not necessarily true that all the right choices must be made in a rigid sequential order, and two, that it is often possible to make functionally equivalent structures by following radically different paths, as shown by the existence of the octopus eye. Indeed, mollusks exhibit a number of different eye structures. Based on these considerations, a realistic estimate of the number of generations needed to develop the eye~\cite{nilsson} yields a value of about 300\,000 generations, a large number, but which requires a few hundred thousand years in a history of life that is estimated to have lasted billions of years.

\subsubsection*{Evolution: The ladder and the tree.} 
\begin{figure}[htb]
\begin{center}
\includegraphics[width=0.7\textwidth]{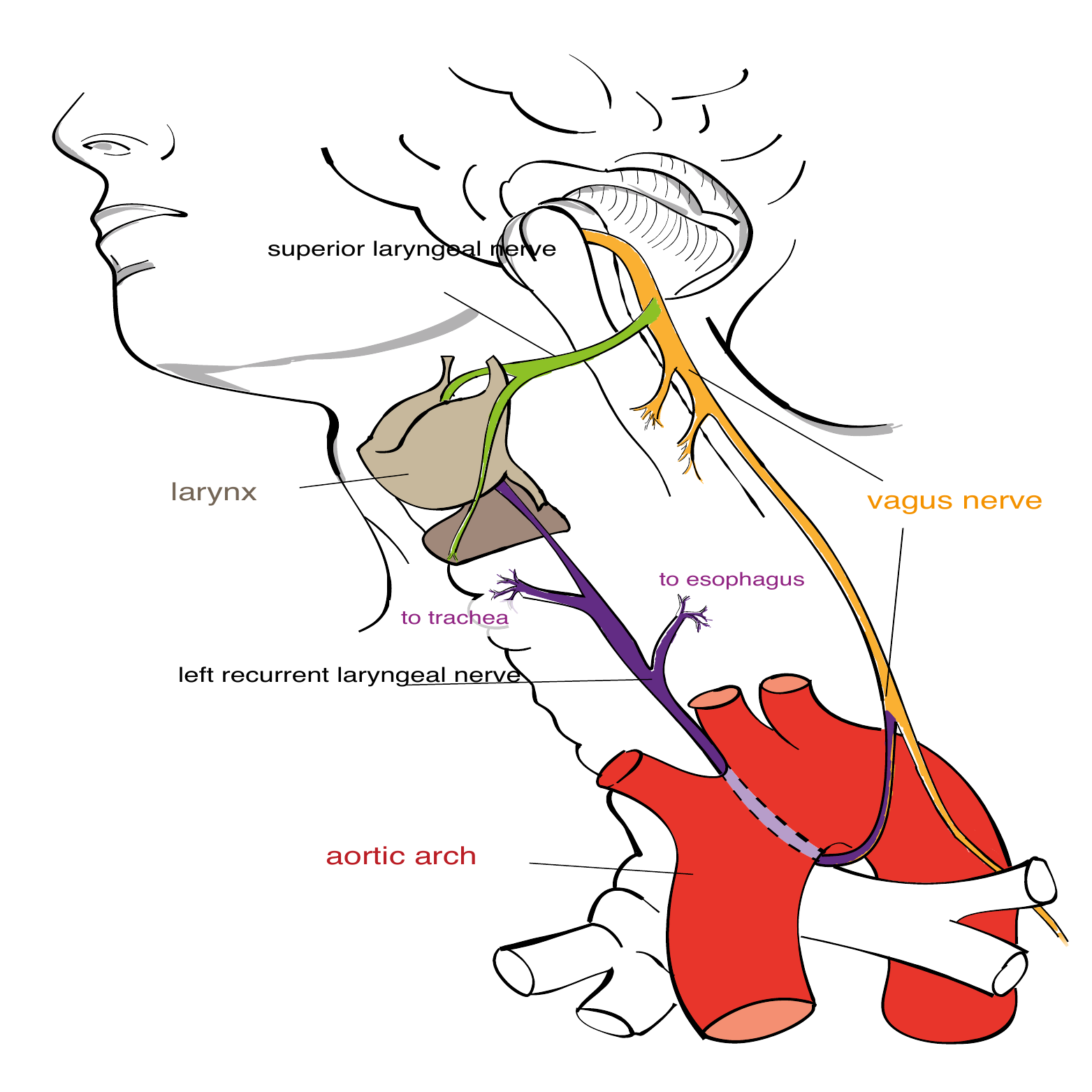}
\end{center}
\caption{The left recurrent laryngeal nerve is a branch of the vagus nerve, which connects with almost all intrinsic muscles of the larynx. Its name derives from the recurrent path it performs with a descent to the thorax, followed by a rise upwards, in the neck, to the larynx. From~\textsl{Wikipedia}.}
\label{fig:nlr}
\end{figure}

In fact the path of evolution is not defined a priori but is an outcome of the evolutionary process itself. As stressed by Jacob~\cite{Jacob},
\begin{quote}\small
Natural selection has no analogy with any aspect of human behavior. However, if one wanted to play with a comparison, one would say that natural selection does not work as an engineer works. It works like a tinkerer---a tinkerer that does not know exactly what he is going to produce but uses whatever he finds around him whether it be pieces of string, fragments of wood or old cardboards: in short it works like a tinkerer who uses everything at his disposal to produce some kind of workable object.
\end{quote}
We can identify traces of this tinkering process by paying attention to those small revealing imperfections where some structures in living organisms appear excessively contrived with respect to what an engineer could have planned from scratch. A well-known example of this structure in mammals (and indeed, in all tetrapods) is the left recurring laryngeal nerve (cf.~fig.~\ref{fig:nlr}), which performs a long detour under the aortic arch to end up connecting to structures under the larynx a few inches away from its branching point form the vagus nerve.  The detour amounts to about 15$'$ in giraffes! 
\begin{figure}[htb]
\begin{center}
\includegraphics[width=\textwidth]{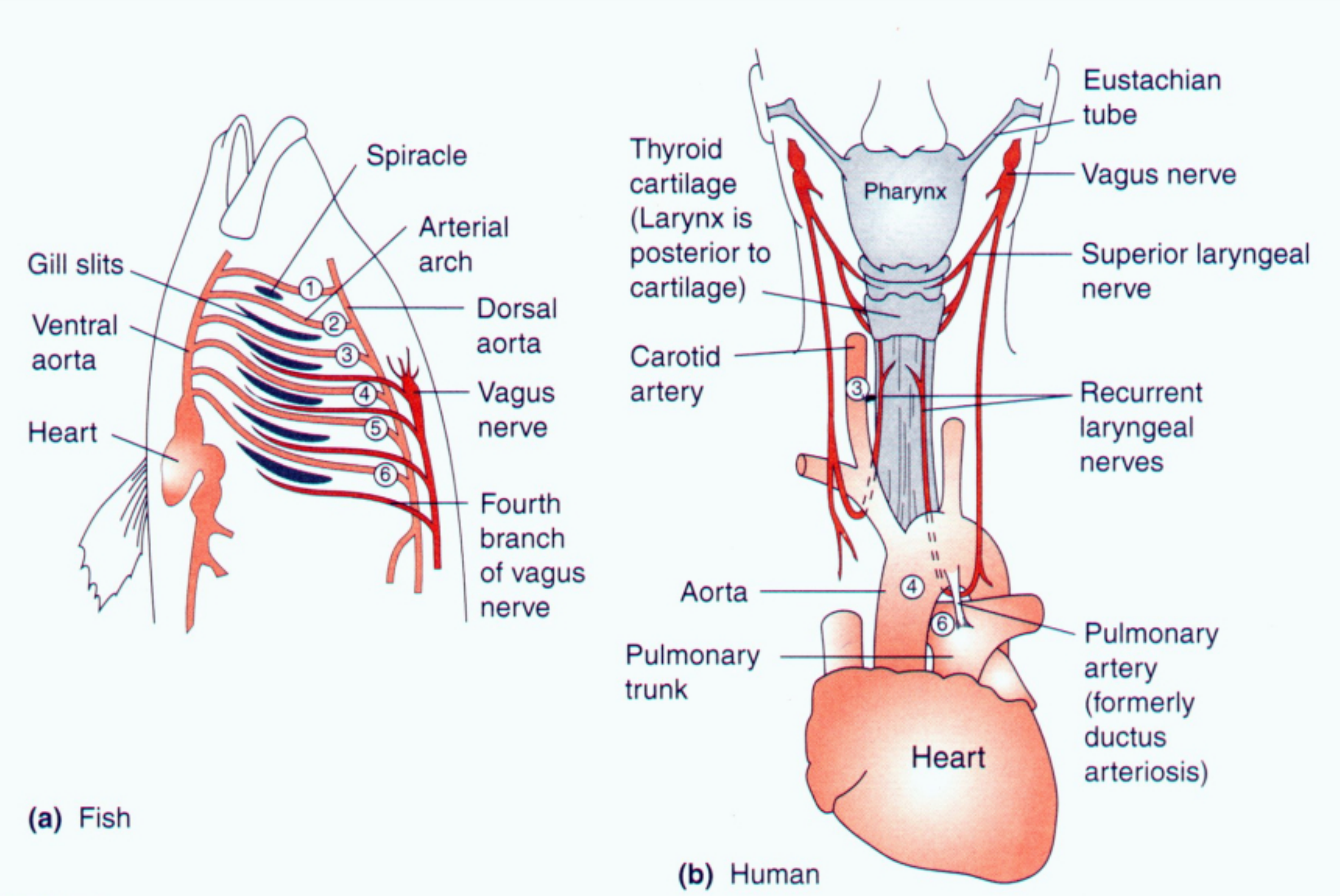}
\end{center}
\caption{Scheme of the evolution of the recurrent laryngeal nerve, from fishes to mammals. The ancestor of this nerve passed close to the arterial arch marked with 6 in the image. This arch has become the ductus arteriosus in mammals. The branches between arch 3 and 4 and between arch 4 and the last one are no more present in mammals. From~\cite[p.~45]{Strickberger}}
\label{fig:nlr2}
\end{figure}
Comparative anatomy allows us to explain this odd fact. As we see in figure~\ref{fig:nlr2}, one of the nerves connecting the fish brain to the gills has become the laryngeal nerve. In the fish this nerve lies behind a neighboring arterial arch (number 6 in the figure). During evolution this arch, that has become the ductus arteriosus, has kept a position close to the heart, and therefore the nerve, which passes behind it and could not cross the arch has been forced to grow longer and longer. 
This has even more striking implications, as argued in~\cite{Wedel}:
\begin{quote}\small
The recurrent course of the nerve from the brain, around the great vessels, to the larynx, is shared by all extant tetrapods. Therefore we may infer that the recurrent laryngeal nerve was present in extinct tetrapods, had the same developmental origin, and followed the same course. The longest-necked animals of all time were the extinct sauropod dinosaurs, some of which had necks 14 meters long. In these animals, the neurons that comprised the recurrent laryngeal nerve were at least 28 meters long. 
\end{quote}

Evolving life forms must modify their structures along the way, without preventing the working of the organism. This is as paradoxical as trying to change an aircraft's propellers during the flight. Therefore, in the process, some less relevant aspects can be neglected with respect to more urgent ones: thus in the giraffe, the cost of lengthening the nerve a few millimeters at each time can be neglected with respect to the advantage of having a longer neck, what allows the animal to reach the leaves of higher trees. 

As any DIY fan knows well, the pleasure of tinkering does not lie in building up things according to instructions, as when assembling IKEA$^{\circledR}$ furniture, but in inventing new usages for the objects in one's possession. In some cases, these object solve, possibly in an original way, already present problems: thus one can use old telephone books to build a stool. In other cases, an unexpected combination of tools allows one to build up an instrument that did not exist before:  possibly the best known example is the way in which Gutenberg made use of the wine press to press paper sheets agains the types. Working as a tinkerer, evolution shows some similar phenomena. Here is how Jacob~\cite{Jacob} summarizes the theory of lung evolution in terrestrial vertebrates, according to Mayr~\cite{Mayr}:
\begin{quotation}\small
Its development started in certain freshwater fishes living in stagnant pools with insufficient oxygen. They adopted the habit of swallowing air and absorbing oxygen through the walls of their esophagus. Under these conditions, enlargement of the surface area of the esophagus provided a selective advantage. Diverticula of the esophagus appeared and, under continuous selective pressure, enlarged into lungs. Further selection of the lung was merely an elaboration of this theme---enlarging the oxygen uptake and vascularization. To make a lung with a piece of esophagus sounds very much like tinkering.
\end{quotation}
This shows how the tinkering evolution can lead by trial and error to a major innovation, creating an organ that did not exist before, like the lung.

Coming back to the example of the garden of forking paths, we were perhaps too pessimistic in assuming that a single wrong path choice at a fork would necessarily lead to ruin. Since there are so many combinatorial possibilities in living life forms, it may well be the case that the path leads---after a great deal of branching---to a different working structure, possibly with a different function. And we can realize that this is possible by looking at the prodigious variety of life forms that exist at present. Gould~\cite{Gould} remarks that 
\begin{quotation}\small
The model of the ladder is much more than merely wrong. It never could provide the promised illustration of evolution progressive and triumphant -- for \textit{it could only be applied to unsuccessful lineages.} 
\end{quotation}
This is the case of the equine family, which exhibits at present no more than six species (of which only one of horses), and of the \textsl{Hominidae} family, the great apes, which contains eight species: three of orangutans, two of chimps, two of gorillas, and \textsl{Homo sapiens}. Many other families cannot fit into a linear scheme. Among the mammals, the \textsl{Mustelidae} family (which contains the badger and the ferret) contains at least 57 species, the \textsl{Cricetidae} one 600, and the \textsl{Muridae} family (which contains mice and rats) more than 700. And we can also think at the vast variety of species of insects. A possibly apocryphal story reports that the evolutionary biologist Haldane answered in this way a theologian who had asked him what he had figured out about the Creator by investigating the creation: 
\begin{quote}\small
The Creator, if He exists, has an ``inordinate fondness for beetles''. 
\end{quote}
Haldane was referring to the fact that there are more than 300\,000 species of beetles against no more than 9000 species of birds and slightly more than 10\,000 species of mammals. Thus, if it may make some sense to look at the evolution of the horse starting from the arrival point (but it is a bit arbitrary to start from the horse, rather than from the zebra or the humble donkey) it will make much less sense to describe the evolution of beetles starting form a randomly chosen present species. 

\begin{figure}[htb]
\begin{center}
\includegraphics[height=0.36\textheight]{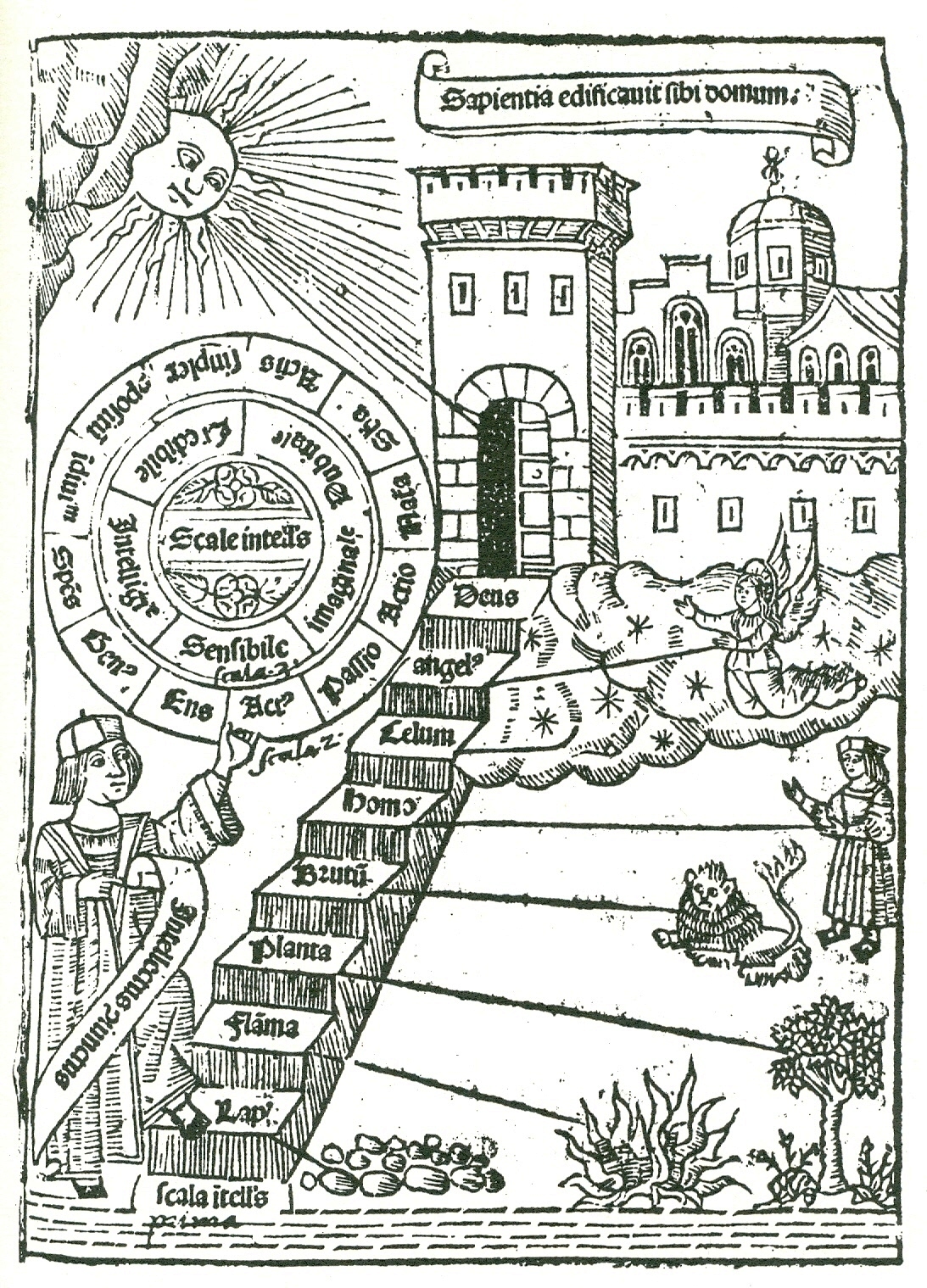}\hspace{0.5cm}\includegraphics[height=0.35\textheight]{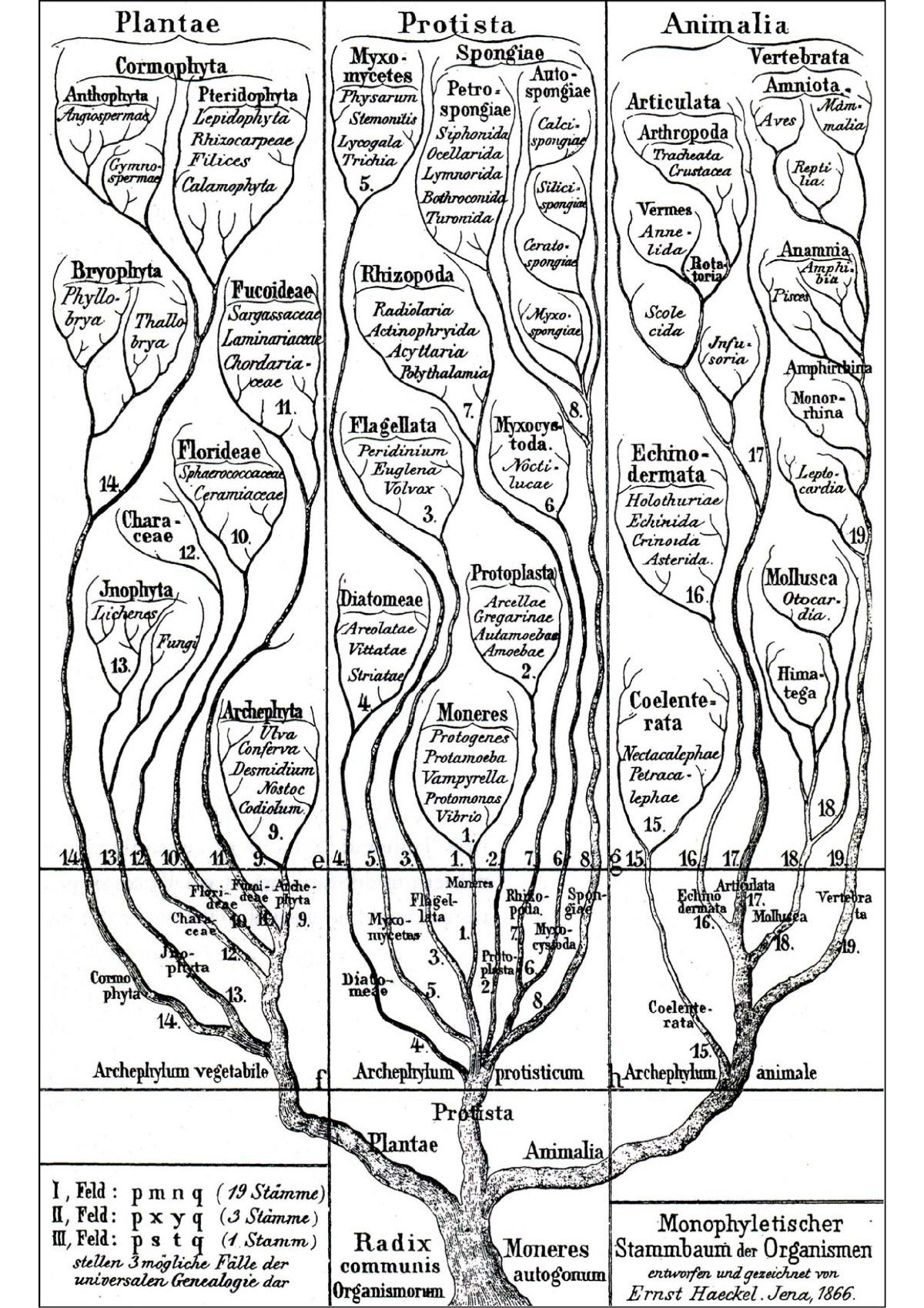}
\end{center}
\caption{The ladder and the tree. Left: The \textit{Scala natur\ae} as represented in R. Lull, \textsl{Liber de ascensu et descensu intellectus} (1305) (Valencia: Jorge Costilla, 1512). Notice that the lower steps correspond to stones (\textsl{Lapis}) and fire (\textsl{Flama}). Right: Monophyletic genealogical tree of the organisms, in E. Haeckel, \textsl {Generelle Morphologie der Organismen} (Berlin: Georg Reimer, 1866).}
\label{fig:llull}
\end{figure}

In normal cases, where a family experiences success in evolution, the representation should make evident the progressive diversification of life forms. The expression used by Darwin to describe his theory was ``descent with modifications''.\footnote{Indeed, the only figure contained in the \textsl{Origin of Species} is a diagram in Chap.~IV representing the divergence and branching of species.} Tracing back from ancestor to ancestor we realize that different life forms have common ancestors: they became different by taking different paths at some fork. Thus a more faithful representation of the evolutionary process is that of a tree. While a linear representation of evolution strongly recalls the medieval images of the \textit{Scala natur\ae} (like the one shown on the left of~fig.~\ref{fig:llull}), already a few years after the publication of the \textsl{Origin} the German biologist E. Haeckel boldly published a single phylogenetic tree encompassing all life forms (on the right of fig.~\ref{fig:llull}, divided in three big branches, and with a common trunk corresponding to the~\textsl{Monera}, that we can identify with today's \textsl{Bacteria} (a term which did not yet exist at that time).
Haeckel introduced, besides branches representing the traditional kingdoms of animals and plants, already recognized by Linnaeus in the \textsl{Systema Natur\ae} and the only ones mentioned by Darwin in the \textsl{Origin}, a third branch, named \textsl{Protista}, which contained all the microscopic organisms known at the time, apart from the \textsl{Monera}.

\begin{figure}[htb]
\begin{center}
\includegraphics[height=0.8\textwidth,angle=270]{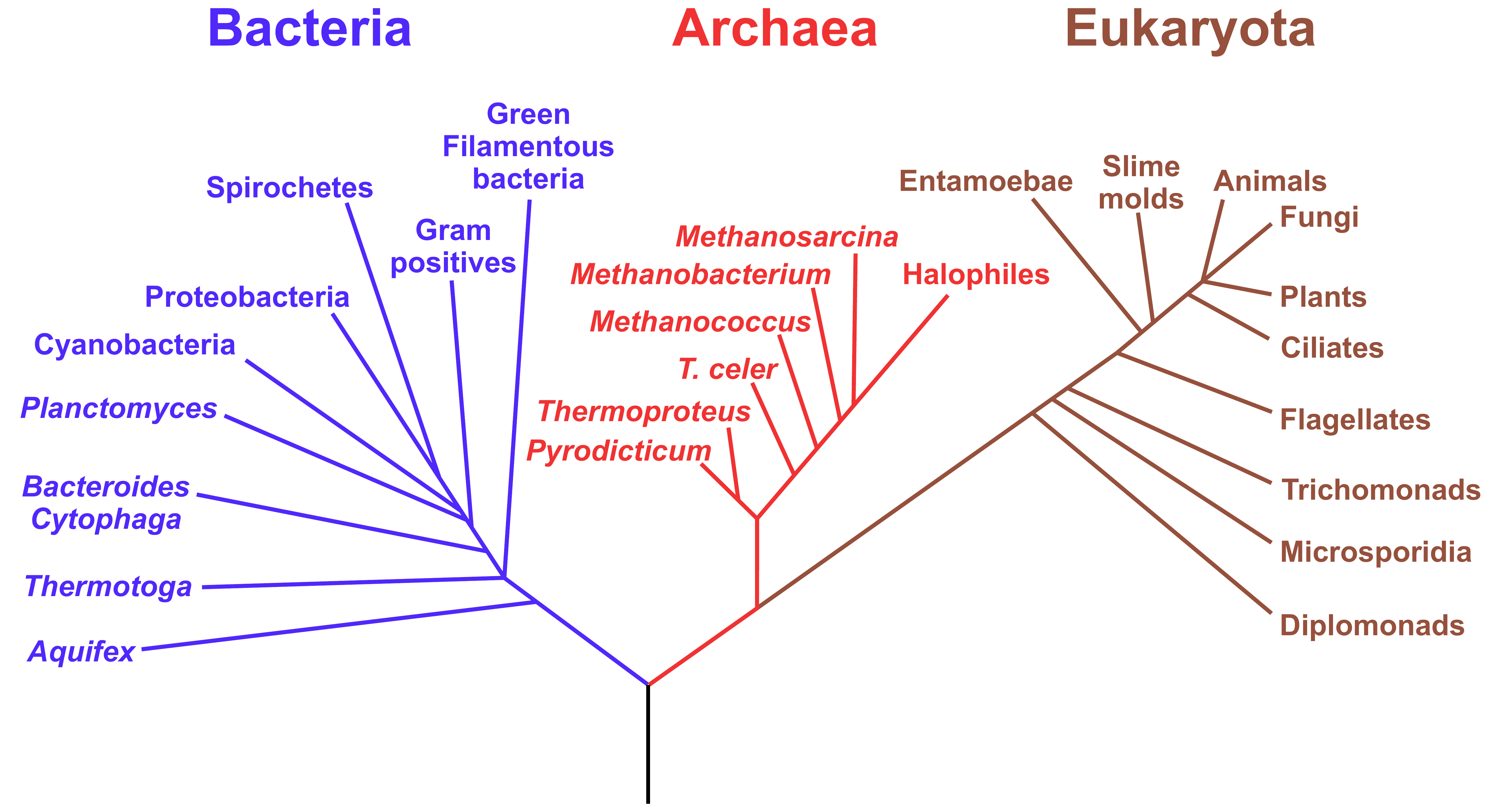}
\end{center}
\caption{A modern schematic view of the phylogenetic tree of living forms. From \textsl{Wikipedia}.}
\label{fig:Tree}
\end{figure}

Haeckel's tree (at least in this edition, since a few years later he moved to an image much closer to the \textit{Scala natur\ae}, with a unique powerful trunk culminating in the man) looks quite close to the one presently accepted by evolutionary biologists, and that is shown, strongly simplified, in figure~\ref{fig:Tree}.
We can identify three main branches, but if we pay a closer attention we see that animals and plants make together with the \textsl{Fungi} (the proper mushrooms) just some terminal twigs of the same branch. The three branches now correspond to  \textsl{Eukaryota}, namely organisms made of cells which possess a well-identified nucleus, and which encompass all multicellular organisms and many unicellular ones; \textsl{Bacteria}, unicellular organisms, which also encompass most of the agents causing infectious diseases (viruses are not considered living beings in this classification); and \textsl{Archaea}, a large group of unicellular organisms that differ from all other life forms by some fundamental aspects of the cell membranes and of the mechanisms of protein synthesis. Many archaea are extremophiles, that is they can survive at high temperatures and pressures, where ordinary bacteria do not survive. It is thought therefore that they were better suited to thrive in the environment of the young Earth, and that therefore they are older than the bacteria, what justifies calling them  \textit{Archaea}. Although the existence of this branch was not suspected as recently as about 50 years ago, it is now believed to be one of the major actors in evolution. According to current opinion, as shown in the figure, the eukaryots branched from the archaea rather than from the bacteria. It has been suggested that the eukaryots originated from cases in which some archaeas (very different from their present descendants) formed symbiotic structures with bacteria that they had engulfed. This stresses how many surprises have been found, also recently, in the study of the fundamental phylogenetic organization of life. 

We can summarize this discussion by stressing two main points: one, that an evolutionary history that appears extremely improbable a priori looks much more probable a posteriori, looking back from the arrival point, since a constant selective pressure has acted during the whole path, favoring advantageous variants with respect to disadvantageous ones. Stated otherwise, the a posteriori probability of a given evolutionary path is a conditional probability, \textit{conditioned by the fact that the concerned life form is presently alive.} It is thus very different from the a priori probability. The second point is that one should not forget that the organisms are not constrained to follow a preexistent path, but can follow a great number of viable pahts, as shown by the great variety of life forms. 

\subsubsection*{The random nature of natural selection.} 
\begin{figure}[htb]
\begin{center}
\includegraphics[width=0.5\textwidth,angle=90]{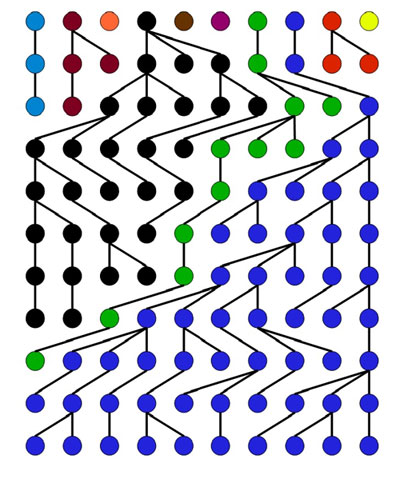}
\end{center}
\caption{Scheme of the Wright-Fisher model of the evolution of an asexually reproducing population. Columns correspond to successive generations of a population of 10 individuals. Different phenotypes are denoted by different colors. One assumes that each individual carries the same phenotype as its parent. Each individual is connected to its parent (in the previous generation) by a black line.}
\label{fig:WF}
\end{figure}

A number of probabilistic techniques have been developed since the thirties of last century to understand the details of the evolutionary process. They form the discipline known as Population Genetics. The best known exponents of this discipline are  R.~A.~Fisher (British), S.~Wright (American) and~M.~Kimura (Japanese). One can consider, for instance, a population formed by a fixed number of individuals which evolves according to the Darwinian mechanism of reproduction, selection and mutation. The simplest model that describes this situation is called the Wright-Fisher model, and is schematically represented in figure~\ref{fig:WF}. 
Columns represent the composition of the population in successive generations, from left to right. The colors represent the features (phenotypes) of the individuals that form the population: different features correspond to different colors. The fitness of a phenotype is measured by the average number of fecund individuals that an individual possessing that phenotype can produce in the given conditions. One assumes an asexual reproduction mechanism and that the phenotype of the offspring is equal to that of its parent, except for mutations (in the represented case there appear no mutations). The new generation is obtained by choosing at random a parent in the previous generation (with probability proportional to its fitness). The parent-child relationship is represented by a black line in figure~\ref{fig:WF}. Due to inevitable fluctuations, some individuals (like the yellow or the magenta ones in the first column) will have no children, while others (like the redo or the black one) will have more than one. (In the represented case, all phenotypes have the same fitness.) Iterating the process, we see that some phenotypes tend to invade the population: in the figure, this happens before with the black phenotype, and eventually with the blue one, which covers the whole population starting from the tenth generation. Kimura has stressed that this phenomenon (called fixation of a trait) can take place in populations made of a small number of individuals even if the phenotypes are equivalent from the selection viewpoint. There is even a small but non-vanishing probability that a sub-optimal phenotype undergoes fixation. Since many populations in nature are made of small numbers of individuals, Kimura suggested that a large part of the diversity of life forms is due to the random fixation of different but selectively equivalent variants in different populations. \begin{figure}[htb]
\begin{center}
\includegraphics[width=0.7\textwidth]{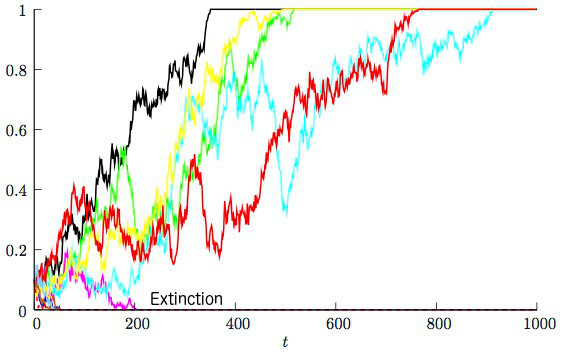}
\end{center}
\caption{Evolution of the frequency of the optimal phenotype in a Wright-Fisher model. Population of 500 individuals. The process has been repeated 10 times, starting from the same initial condition. The optimal phenotype has a 1\% larger fitness than the others. The initial frequency of the optimal phenotype is 10\%. The optimal phenotype fixates in 5 out of 10 cases, and goes to extinction in the other cases.}
\label{fig:WF4}
\end{figure}
Figure~\ref{fig:WF4} represents the evolution of the frequency of the optimal phenotype in a population of 500 individuals. The fitness of the optimal phenotype is larger by 1\% with respect to the others, meaning that an individual that carries it has 1\% more offspring (on average) than one that doesn't. The initial fraction of individuals carrying the optimal phenotype is 10\%. In this situation we observe that the optimal phenotype goes to fixation within 1000 generations in 5 cases out of 10, and gets extinguished in the other cases. Thus natural selection has a positive effect for this case only 50\% of the times, due to the small population size. Actually one of the most relevant results obtained by Kimura is that if in a population there is a single mutant having a fitness advantage, say, of 1\% with respect to the remainder of the population, the probability that the trait eventually goes to fixation is equal to its relative advantage, that is, in our case, is of 1\%. (Or even, taking into account that in sexually reproducing populations each individual has two copies of the genome, and the mutation appears on only one of them, this probability will be only half of 1\%.)

\subsubsection*{Is evolution predictable?} Given that sometimes even natural selection is not effective, what hope can we have to predict evolutionary processes? Or should we be contented to use it to explain a posteriori the properties of living beings, without daring to frame predictions? 

Actually predictions based on the theory of evolution have been stated and verified many times. In a  very interesting case, first Darwin and later (independently) the other discoverer of natural selection, A. J. Wallace, made a precise prediction. This prediction was verified with complete satisfaction only many years later and the conclusion was reached a few years ago, in 1992, with the discovery of the ``smoking gun''. This is a story worth telling~\cite{Arditti}. As we know, pollination plays an essential role in the life cycle of flowering plants, and they often produce nectar to lure insects into visiting their flowers. It is clearly advantageous for the plant to develop mechanisms that help the visit of a particular species of insects rather than another, because this increases the probability that their precious pollen is carried to flowers of the same species. Under the pressure of evolution, many flowering species have produced elaborate ways to discriminate among the different species of insects that collect their nectar. Darwin dedicated one of his treaties to ``The Various Contrivances By Which British And Foreign Orchids Are Fertilised By Insects''. Knowing of Darwin's interest in orchids, an orchid grower sent him in 1862 a box containing in particular the beautiful star-shaped flower of~\textsl{Angraecum sesquipedale}. The name is an allusion to the fact that the nectar is located at the bottom of a tube almost a foot long. In a letter to a friend, Darwin wrote ``Good Heavens what insect can suck it’' and in another letter a few days later he suggested that ``in Madagascar there must be moths with probosces capable of extension to a length of between ten and eleven inches [25.4–27.9 cm]''. Darwin published this prediction in his book on orchids, published the same year. In 1907, more than 20 years after Darwin's death, a subspecies of the moth~\textsl{Xantopan morganii}, already remarked in Congo by its long proboscis, was identified in Madagascar. This moth has a wingspan of about 16cm, but its proboscis is truly colossal, more than 20cm long, and it forms a large coil before the head when not in use. Given Darwin's prediction, this subspecies was called~\textsl{Xantopan morganii praedicta}.
\begin{figure}[htb]
\begin{center}
\includegraphics[width=0.77\textwidth]{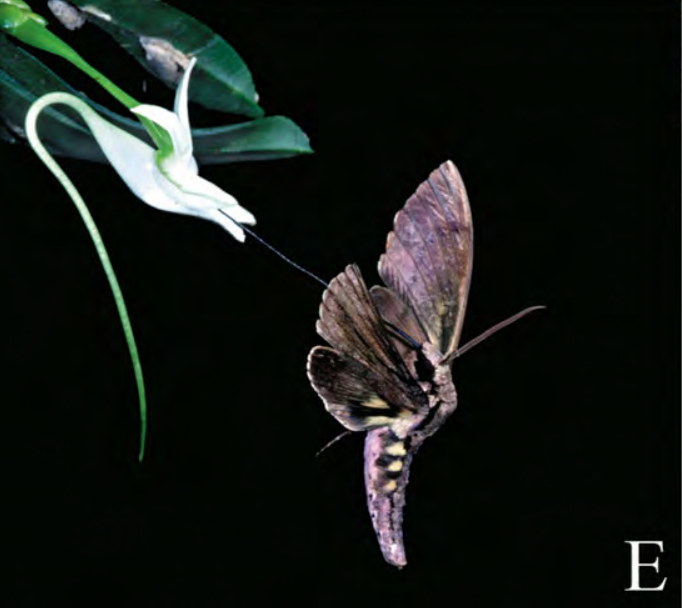}
\end{center}
\caption{Caught in the act! \textsl{Xantopan morganii praedicta} sucks \textsl{Angraecum sesquipedale}'s nectar. From~\cite{Arditti}.}
\label{fig:xantopan}
\end{figure}
We had however to wait until 1992, about 130 years after Darwin's prediction, to be able to observe moths of this subspecies in the act of sucking nectar from \textsl{Angraecum}'s flower and of carrying its pollen from plant to plant (cf.~fig.~\ref{fig:xantopan}). \textsl{Angraecum sesquipedale} is now known as ``Darwin's orchid'' among the aficionados.

In the recent years there has been a growing effort to go from reconstruction and analysis of the evolutionary processes to the prediction of their future behavior. The interest is not only academic. We can produce an effective influenza vaccine only if we have reliable predictions on the evolutionary dynamics of its virus. Indeed, the evolution of influenza virus exhibits two apparently contradictory features: on the one hand its mutation rate is so high that the dominating strain changes substantially every few years, and on the other hand in each epidemic season most infections are due to closely related strains. This second feature allows the production of an effective vaccine, provided the dominant strain has been identified with an advance long enough to allow the production of enough doses. A group of the World Health Organization (WHO) is dedicated to the identification of the prevailing strain for each coming epidemic season. It exploits world-wide collected data on the prevalence of the different viral strains and on the effectiveness on the different strains of the immunity distributed in the population due to the previous epidemics.~\cite{Influenza} Every six months, the WHO publishes its recommendations for the vaccine composition, about six months before the beginning of the epidemic season for each hemisphere. 

How can one make predictions for an evolutionary process, when it is determined by so many random factors? When only few data were available, mostly referring to macroevolutionary processes (species evolution), it was natural to be skeptical about the possibility of repeating and therefore of predicting such processes. But things are now changing, since we can now have high throughput data on sequences and on phenotypes, we can perform experiments in parallel on many samples, and we are developing more effective methods to analyze the behavior of complex dynamical systems. This opens the way to the possibility of formulating and validating predictions about evolutionary processes on an accessible time scale: and this could help us to solve important problems like the selection of the influenza vaccine, but also to fight against the development of antibiotic resistance (especially in hospital environment) or in the prediction of cancer development in a single patient: since cancer evolution reproduces at a small scale and in a shorter time span a process of Darwinian selection.

\begin{figure}[htb]
\begin{center}
\includegraphics[width=0.5\textwidth]{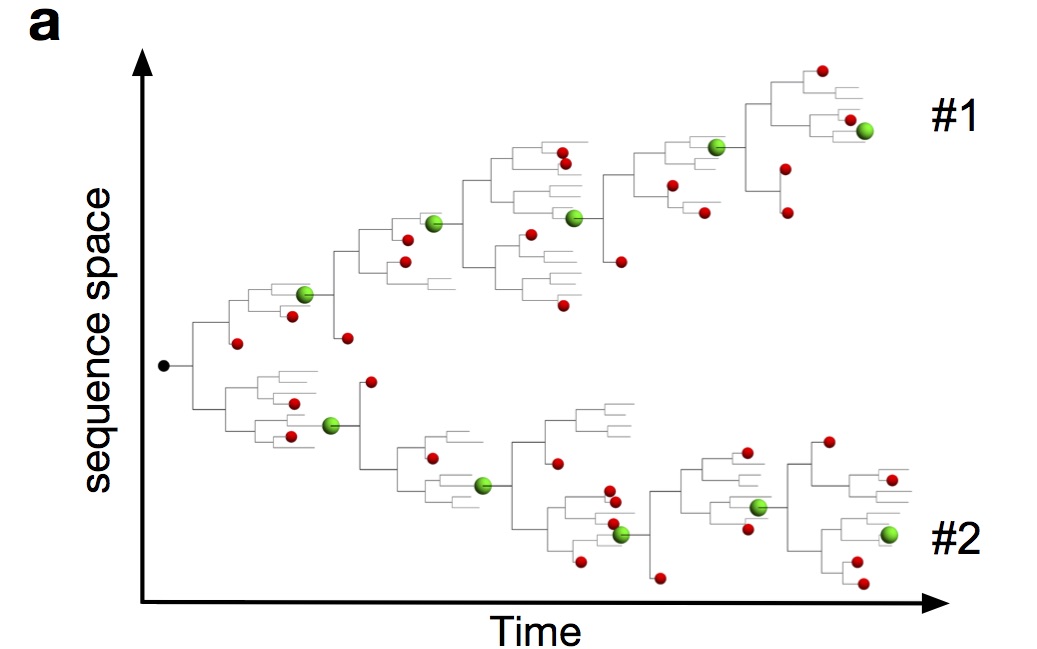}\includegraphics[width=0.5\textwidth]{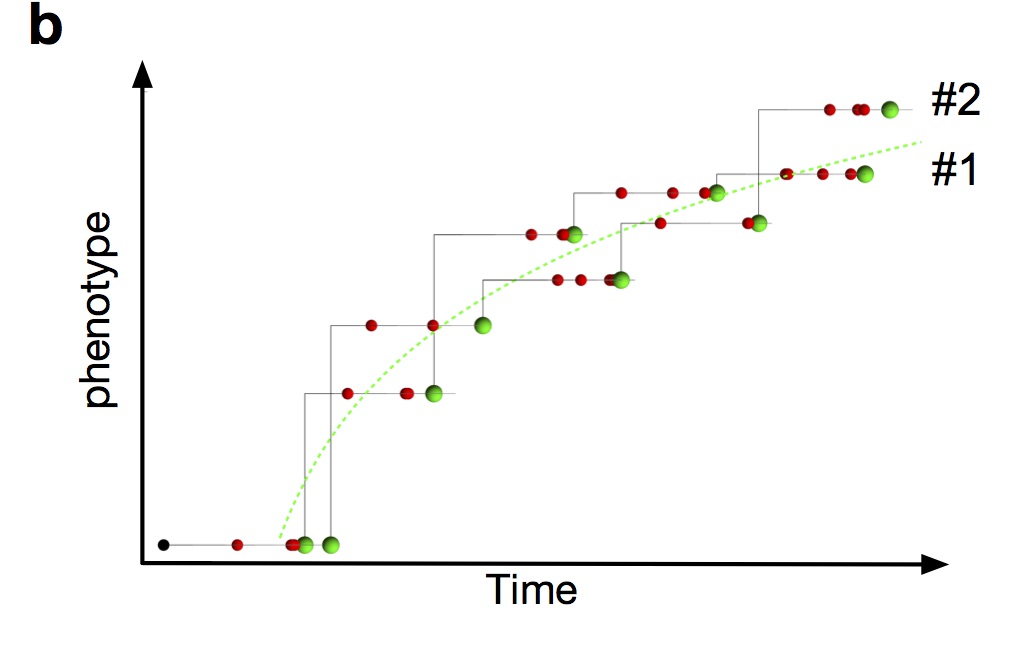}
\end{center}
\caption{Schematic representation of a selection-mutation process in genotypic and in phenotypic space. Comparison of the evolutionary process in two parallel populations in two different representations. Left: Evolution is sequence (genotype) space is a stochastic process with many equiprobable pathways. Mutations produce new variants, represented as nodes in a tree. These variants are only weakly checked by selection. Those positive selected (advantageous) are represented by green dots, and those negatively selected (disadvantageous) by red ones. The two populations follow diverging paths.  Right: Fitness and other phenotypic traits evolve in a regular way. The pathways shown contain hte same mutations (green and red) as on the left panel, but now different sequence produce similar phenotypic effects. Positive and negative selecion effects lead the evolution towards a single phenotypic path, represented by the green line. The populations evolve in parallel close to this path. From~\cite{MWL}.}
\label{fig:scheme2}
\end{figure}
While it will probably be always impossible to predict the evolution of the \textit{genotype} of life forms, it could be possible to attain a less ambitious goal, namely to predict---within limits---the evolution of their \textit{phenotypes}. The genotype of an individual is the whole of its inheritable characteristics: in practice, it is represented for a good part by its genome, that is by its DNA sequence. As we know the DNA is described by a very long sequence written in a four-letter alphabet:  \texttt{A}, \texttt{T}, \texttt{C} and \texttt{G}. In the simplest organisms, this sequence is already a few million letter long, and the number of possible variants is enormous. Evan if we consider just those variants with only a few differences from a given sequence, the number of possible variants is already of the order of the genome length. Many of these variants are irrelevant, because they do not modify the phenotype of the organism, that is the whole of its \textit{observable} inheritable characteristics. But the observable features are the ones that fall under the edge of natural selection. This point is schematically represented in figure~\ref{fig:scheme2}, which has been published in a recent paper on the predictability of evolution. An experiment made by a group in Harvard on the common beer yeast \textsl{Saccaromyces cerevisiae}~\cite{Desai} has shown that this is not just a theoretical construction. 
\begin{figure}[htb]
\begin{center}
\includegraphics[width=0.7\textwidth]{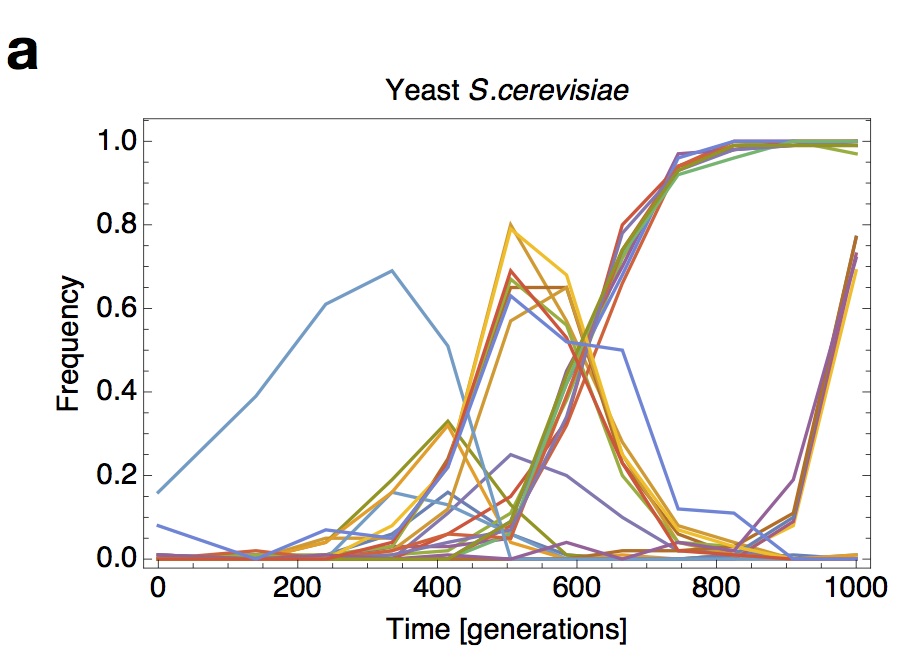}
\end{center}
\caption{Frequency of several variants in \textsl{Saccaromyces cerevisiae} populations evolving in parallel. Although the variants are genotypically different, one can identify some similarities in their dynamics. Typically some variants first grow in frequency, but are later overcome by newly arising ones. This shows that parallel evolution exhibits some degree of predictability. From~\cite{MWL}, on data by~\cite{Desai}.}
\end{figure}
In this experiment 640 yeast strains have been allowed to evolve in parallel for 1000 generations under identical conditions. Their fitness was measured from time to time by measuring their growth rate and comparing it to that of the ancestral strain. Thus the experimenters could evaluate the fitness variations among the different strains, and how the fitness of a strain depended on the fitness of its ancestor. Since natural selection acts on the phenotype it is quite reasonable that evolutionary dynamics is more predictable at the phenotypic rather than at the genotypic level. Exploiting this concept it has been possible to develop a predictive model~\cite{LL} of the evolution of the influenza virus A/H3N2 which should allow to improve on the WHO method, reaching a predictability horizon of the order of one year. 

One of the founders of the evolutionary synthesis, Dobzhansky, famously stressed that ``Nothing in biology makes sense except in the light of evolution''.~\cite{Dobzhansky} We can add to this that evolution cannot be understood except in the light of probability theory. And that the tools of probability theory can already shed light on evolutionary processes that touch us directly and hopefully will help us to control them in the near future. 


\begin{thebibliography}{99}
\bibitem{Hoyle}Hoyle, F. and Wickramasinghe, N. C., \textsl{Evolution from Space: A Theory of Cosmic Creationism} (New York: Simon and Schuster, 1981). \bibitem{Dawkins}Dawkins, R., \textsl{The Blind Watchmaker} (London: Penguin Books, 1986). 
\bibitem{nilsson}Nilsson, D.-E. and Pelger, S., A pessimistic estimate of the time required for an eye to evolve, \textit{Proc. R. Soc. Lond. B} \textbf{256} 53-58 (1994)
\bibitem{Jacob}Jacob, F., Evolution and tinkering, \textit{Science} \textbf{196} 1161-1166 (1977)
\bibitem{Wedel}M. J. Wedel, A monument of inefficiency: The presumed course
of the recurrent laryngeal nerve in sauropod dinosaurs, \textit{Acta Palaeontologica Polonica} \textbf{57} 251-256 (2012)
\bibitem{Strickberger}M. W. Strickberger, \textsl{Evolution} (3rd ed.)  (Sudbury MA: Jones and Bartlett, 2005)
\bibitem{Mayr}Mayr, E., From molecules to organic diversity, \textit{Federation Proceedings} \textbf{23} 1231-1235 (1964), reprinted in: Mayr, E. \textsl{Evolution and the Diversity of Life} (Cambridge (Mass.): Belknap, 1976) cited by~[4]. For a technical description, see Packard, G. C, The evolution of air-breathing in Paleozoic gnathostome fishes, \textit{Evolution} \textbf{28} 320-325 (1974). More recently, this model has been the target of some criticism. See, e.g.: Farmer, C., Did lungs and the intracardiac shunt evolve to oxygenate the heart in vertebrates? \textit{Paleobiology} \textbf{23} 358-372 (1997) 
\bibitem{Gould}Gould, S. J., Life's little joke, in: Gould, S. J., \textsl{Bully for Brontosaurus: Reflections in Natural History} (New York: Norton, 1991). 
\bibitem{Kimura}Kimura, M., On the probability of fixation of mutant genes in a population, \textit{Genetics} \textbf{47} 713-719 (1962)
\bibitem{Arditti}Arditti, J., Elliott, J. Kitching, I. J., and Wasserthal, L. T., `Good Heavens what insect can suck it' -- Charles Darwin, \textsl{Angraecum sesquipedale} and \textsl{Xantopan morganii praedicta}, \textit{Botanical Journal of the Linnean Society} \textbf{169} 403-432 (2012)
\bibitem{Influenza}One can find the description of the protocul used by the WHO at the URL~\url{http://apps.who.int/gb/pip/pdf_files/Fluvaccvirusselection.pdf?ua=1}
\bibitem{MWL}Lässig, M., Mustonen, V., and Walczak, A. M., Predicting evolution, \textit{Nature Ecology \&\ Evolution} \textbf{1} 0077 (2017)
\bibitem{Desai}Kryazhimsky, S., Rice, D. P., Jerison, E. R., and Desai, M. M., Global Epistasis Makes Adaptation Predictable Despite Sequence-Level Stochasticity, \textit{Science} \textbf{344} 1519-1522 (2014)
\bibitem{LL}Łuksza, M., and Lässig, M., A predictive fitness model for influenza, \textit{Nature} \textbf{507} 57-61 (2014)
\bibitem{Dobzhansky}Dobzhansky, Th., Nothing in Biology Makes Sense Except in the Light of Evolution, \textit{American Biology Teacher} \textbf{35} 125-129 (1973)
\end{thebibliography}
\end{document}